\documentstyle[12pt]{article}
\newcommand\be{\begin{equation}}
\newcommand\ee{\end{equation}}
\newcommand\ba{\begin{eqnarray}}
\newcommand\ea{\end{eqnarray}}

\newcommand{\cl}{\centerline}
\newcommand\bear{\begin{eqnarray*}}
\newcommand\eear{\end{eqnarray*}}

\def\half{\frac{1}{2}}
\begin{document}
\begin{titlepage}
\setlength{\textwidth}{5.0in}
\setlength{\textheight}{7.5in}
\setlength{\parskip}{0.0in}
\setlength{\baselineskip}{18.2pt}
\hfill
{\tt HD-THEP-04-16}
\begin{center}
{\large{\bf Gauge Identities and the Dirac Conjecture}}\par
\vskip 0.3cm
\end{center}
\begin{center}
{Heinz J. Rothe and Klaus D. Rothe}\par
\vskip 0.3cm
{Institut f\"ur Theoretische Physik}\par
{Universit\"at Heidelberg, Philosophenweg 16, D-69120 Heidelberg, Germany}
\footnote{email: h.rothe@thphys.uni-heidelberg.de\\
k.rothe@thphys.uni-heidelberg.de}
\cl{\today}
\end{center}

\begin{abstract}
\noindent
The gauge symmetries of a general dynamical system can be systematically obtained following either a Hamiltonean or a Lagrangean approach. In the
former case, these symmetries are generated, according to Dirac's
conjecture, by the first class constraints. In the latter approach
such local symmetries are reflected in the existence of so called gauge identities. The connection between the two becomes apparent, if one works with a first order Lagrangean formulation. Our analysis applies to purely first class systems. We show that Dirac's conjecture applies to first class constraints which are generated in a particular
iterative way, regardless of the possible existence of bifurcations
or multiple zeroes of these constraints. We illustrate these statements in terms of several examples.

\bigskip\noindent
PACS: 11:10; 11:15; 11:30
\end{abstract}
\end{titlepage}

%%%%%%%%%%%%%%%%%%%%%%
\section{Introduction}
%%%%%%%%%%%%%%%%%%%%%%%

The problem of revealing the gauge symmetries of a Lagrangean has 
been the subject of numerous investigations \cite{1}-\cite{RR}. 
Local symmetries of a dynamical system have been studied both within the Lagrangean as well as the Hamiltonian framework. 

On the Lagrangean level there exist well known algorithms 
\cite{Sudarshan,Shirzad99,Pyatov,Gomis,RR} 
for detecting the gauge symmetries of a Lagrangean. 
It has the merit of directly generating the 
transformation laws in configuration space, expressed in terms of an
independent set of arbitrary functions, which leave the action invariant. Every one of the
gauge parameters parametrizing such a local symmetry 
is directly related to a so-called ``gauge identity".
The number of such parameters is equal to the number of independent gauge identities.

On the Hamiltonian 
level the relevant action whose vanishing variation leads to the Hamilton 
equations of motion, is the so called ``total action". The transformation laws in phase space, which leave this action invariant, have
been conjectured by Dirac to be generated by the first class constraints 
\cite{Dirac}.
The number of such constraints is in general larger than the number of
gauge-identities of the Lagrangean formulation, referred to above. Hence the number of free functions parametrizing the symmetry of the total action is in
general less or equal to the number of first class constraints. This is a well known fact \cite{Girotti,Henneaux,BRR}. 

In this paper we shall restrict ourselves to purely first class
systems. Starting from an equivalent first order Lagranean
formulation, whose Euler-Lagrange equations are just the Hamilton equations of motion, we make use of well known Lagrangean methods, in order to establish a direct connection between the gauge identities and the generators
of gauge symmetries in the Hamiltonian formulation. In particular we show, that for the first class constraints
to be generators of local symmetries of the total action,  these must be generated in a {\it particular iterative} way. It is in this sense that Dirac's conjecture holds. This generalizes
the results of one of the authors, ref. \cite{Heinz}, to an arbitrary number of primaries. We use this framework in order discuss several examples
which have been cited in the literature as counterexamples to Dirac's conjecture,
in order to point out some subtleties, and to confirm Dirac's conjecture in the sense described above. This is the main objective of the paper.

The organization of the paper is as follows. In section 2 we discuss the iterative generation of the first class chains terminating in gauge
identities for an arbitrary number of primaries, following the general ideas of ref. \cite{Heinz}. We then derive from the gauge
identities the transformation laws for the coordinates, momenta and
Lagrange multipliers, which leave the total action invariant, and show that they have the form conjectured by Dirac, if the first class constraints are
generated in a definite iterative way.
We illustrate in section 3 the formalism for the case of an arbitrary system
with two primary constraints and one secondary constraint. Section 4 is devoted to three concrete examples which have been considered in the literature as counter examples to the Dirac conjecture: the first one discusses the possibility of a first class constraint
becoming ``ineffective" on the level of the equations of motion, though generating an off-shell symmetry on the level of the total action; it also demonstrates that the replacement of a first class constraint generated iteratively by an equivalent one is in clash with Dirac's conjecture. The 
second example illustrates the possibility that our algorithm generates 
the correct local symmetries of the total action, away from the constrained 
surface, eventhough the ``structure functions" may be singular on that 
surface. The third example involves bifurcations of an iteratively generated constraint. We show in this case that the  choice of a particular bifurcation will not correspond to an off-shell symmetry of the total action, while the strict implementation of our algorithm uncovers the full local symmetry of the total action.  
We conclude in section 5, where we once more summarize our general statement regarding Dirac's conjecture.

%%%%%%%%%%%%%%%%%%%%%%%%%%%%%%%%%%%%%%%%%%%%%%%%%
\section{Gauge identities and Dirac's conjecture}
%%%%%%%%%%%%%%%%%%%%%%%%%%%%%%%%%%%%%%%%%%%%%%%%%
The following analysis applies to purely first class systems, and is a extension
of ref. \cite{Heinz} to the case of an arbitrary number of primaries
%%%%%%
\footnote{When refering to that reference, the reader should be aware of a number of notational differences.}
%%%%%%
. 
Let $\phi^{(0)}_{a}=0$ ($a = 1,2,\cdot\cdot\cdot,N$) be 
the primary constraints associated with a {\it second} order Lagrangean $L(q,\dot q)$,
where $q_i,i=1,\cdots,n$ are coordinates in configuration space, and
let $H(q,p)$ be the corresponding Hamiltonian defined on the primary surface. 
One readily verifies that the Euler-Lagrange equations 
associated with the first order (total) Lagrangean
\be\label{Ltotal} 
L_T(q,p,\dot q,\dot p,\lambda,\dot\lambda) = \sum^n_{i=1}p_i\dot q_i 
- H_T(q,p,\lambda)
\ee 
with
\be\label{Htotal}
H_T(q,p,\lambda) = H(q,p) + \sum_{a=1}^N\lambda^{a}\phi^{(0)}_{a}\,,
\ee
reproduces the Hamilton equations of motion including the primary constraints,
if we regard $q_i,p_i$ and $\lambda^{a}$ as coordinates in an 
$2n+N$ dimensional configuration space.
We now write (\ref{Ltotal})
in the form 
\be\label{Ltotal-general}
L_T = \sum^{2n+N}_{\alpha=1} a_\alpha(Q)\dot Q_\alpha - H_T(Q)\,,
\ee
where
\be\label{general-coordinates}
Q_\alpha := (\vec q,\vec p,\lambda^1,\cdot\cdot\cdot,\lambda^{N})
\ee
and
\be\label{Htotal1}
H_T(Q) = H(Q_1,\cdot\cdot\cdot,Q_{2n}) + \sum_{a=1}^N 
Q_{2n+a}\phi^{(0)}_{a}\,.
\ee
The non-vanishing elements of $a_{\alpha}$ are given by $a_{i}=Q_{n+i}=p_i$ 
($i=1\cdot\cdot\cdot,n$).
The $2n+N$ components of the Euler derivative are
%%%%%%%%%
\footnote{Our definition of the Euler derivative differs
from that of ref. \cite{Heinz} by a minus sign. This reference contains some printing errors.
In particular the index ``$a$" in Eq. (56) of that reference takes the values
$a = 2,\cdots,M$, and $\alpha$ in Eq. (59) runs over $1,2,\cdots,2n+1$,
and not from $1$ to $7$, as stated in the paragraph following (59).}
%%%%%%%%%%%%%%%%%
\ba\label{Euler-derivative}
E^{(0)}_\alpha &=& 
\frac{d}{dt}\left(\frac{\partial L_T}{\partial\dot Q_\alpha}\right)
-\frac{\partial L_T}{\partial Q_\alpha}\\
&=& -\sum^{2n+N}_{\beta=1}F^{(0)}_{\alpha\beta}\dot Q_\beta
+ K^{(0)}_\alpha\,,
\ea
 with
\be\label{Falphabeta}
F^{(0)}_{\alpha\beta}= \partial_\alpha a_\beta - \partial_\beta a_\alpha \,.
\ee
$F^{(0)}$ is the $(2n+N)\times(2n+N)$ matrix
\be
{\bf F}^{(0)} = \left(
\begin{array}{ccc}
{\bf 0}&{\bf -1}&\vec 0\cdots \vec 0\\
{\bf 1}&{\bf 0}&\vec 0\cdots \vec 0\\
{\vec 0}^T&{\vec 0}^T&0\cdots 0\\
\cdot&\cdot&\cdot\\
\cdot&\cdot&\cdot\\
\cdot&\cdot&\cdot\\
{\vec 0}^T&{\vec 0}^T&0\cdots 0\\
\end{array}\right)\,,
\ee
and  
\be
K^{(0)}_\alpha = \frac{\partial H_T}{\partial Q_\alpha}\,,
\ee
where ${\bf 1}$ is a $n\times n$ unit martix, $\vec 0$ are $N$-component Null column vectors
(associated with the absence of $\dot\lambda^a$ in $L_T$), and $\vec 0^T$ is the transpose of $\vec 0$.

The variation of the total action
\be\label{Stotal} 
S_T = \int dt\ L_T(Q,\dot Q)
\ee
is given by
\be\label{delta-Stotal}
\delta S_T = -\sum_\alpha\int dt\ E^{(0)}_\alpha\delta Q_\alpha\,,
\ee
where we have dropped a boundary term.
The left-zero modes of ${\bf F}^{(0)}$ are given by
\be
\vec v^{(0)}(a) = \left(\vec 0,\vec 0,{\hat n}(a)\right)\,,
\ee
where ${\hat n}(a)$ is a $N$-component unit vector   
with the only non-vanishing component in the $a$'th  
place. Hence 
\be\label{vdotE0}
\vec v^{(0)}(a)\cdot\vec E^{(0)} = \vec v^{(0)}(a)\cdot\vec K^{(0)} = \phi^{(0)}_{a}\,.
\ee
We thus recover on-shell the (first class) primary constraints
$\phi_a^{(0)} = 0$.

We now adjoin the time derivative of the primaries $\{\phi^{(0)}\}$ 
to $\vec E^{(0)}$ and construct the $2n+2N$ component (level one) vector 
$\vec E^{(1)}$:
\be\label{E1}
{\vec E}^{(1)} = \left(
\begin{array}{c}
\vec E^{(0)}\\
\frac{d}{dt}{\vec\phi}^{(0)}\\
\end{array}\right)\,,
\ee
where $\vec\phi^{(0)}$ is a colum vector with components $(\phi^{(0)}_1,\cdot\cdot\cdot,\phi^{(0)}_{N})$. By construction $\vec E^{(1)}$ vanishes on shell, i.e. for 
$\vec E^{(0)} = 0$. 
The components of ${\vec E}^{(1)}$, which we label by $\alpha_1$, can be 
written in the form
\be
E^{(1)}_{\alpha_1} = -\sum_{\alpha}F^{(1)}_{\alpha_1\alpha}\dot Q_{\alpha} 
+K^{(1)}_{\alpha_1}(Q)\,,
\ee
where 
\be
{\vec K}^{(1)} = \left(
\begin{array}{c}
\vec K^{(0)}\\
\vec 0
\end{array}\right)\,,
\ee
and ${\bf F^{(1)}}$ is now the {\it rectangular} matrix 
\be\label{F1}
{\bf F}^{(1)} = \left(
\begin{array}{ccc}
{\bf 0}&{\bf -1}&\vec 0\cdots\vec 0\\
{\bf 1}&{\bf 0}&\vec 0\cdots\vec 0\\
{\vec 0}^T&{\vec 0}^T&0\cdots 0\\
\cdot&\cdot&\cdot\\
\cdot&\cdot&\cdot\\
\cdot&\cdot&\cdot\\
{\vec 0}^T&{\vec 0}^T&0\cdots 0\\
{-\nabla}\phi^{(0)}_1&{-\tilde\nabla}\phi^{(0)}_1&0\cdots 0\\
\cdot&\cdot&\cdot\\
\cdot&\cdot&\cdot\\
\cdot&\cdot&\cdot\\
{-\nabla}\phi^{(0)}_{N}&{-\tilde\nabla}\phi^{(0)}_{N}&0\cdots 0\\
\end{array}\right)
\ee
Here
\be\label{nabla}
{\nabla} := (\partial_{1},\cdot\cdot\cdot,\partial_{n})\,,\quad
{\tilde\nabla} := (\partial_{n+1},\cdot\cdot\cdot,\partial_{2n})\,.
\ee
We seek new constraints by looking for left zero modes of ${\bf F^{(1)}}$. 
They are $N$ in number, and are given by
\be\label{zeromode1}
\vec v^{(1)}(a) := \left(-{\tilde\nabla}\phi^{(0)}_{a},{\nabla}\phi^{(0)}_{a},
\vec 0,\hat e^{(0)}(a)\right)\,,
\ee
where $\hat e^{(0)}(a)$ is an $N$-component unit vector, with the only non-vanishing component in the $a$'th position. This leads to
\ba\label{level-one}
\vec v^{(1)}(a)\cdot\vec E^{(1)}= \vec v^{(1)}(a)\cdot\vec K^{(1)} 
&=& \frac{\partial\phi^{(0)}_{a}}{\partial q_i}
\frac{\partial H_T}{\partial p_i}
-\frac{\partial H_T}{\partial q_i}
\frac{\partial\phi^{(0)}_{a}}{\partial p_i}\nonumber\\
&=& \{\phi^{(0)}_{a},H_T\}\,,
\ea
or
\be\label{vdotE1}
\vec v^{(1)}(a)\cdot\vec E^{(1)} = \{\phi^{(0)}_{a},H\}  
-\sum_{c} \lambda^{c} \{\phi^{(0)}_{c},\phi^{(0)}_{a}\}\,,
\ee
which by constuction vanish on shell (i.e., for $\vec E^{(0)} = \vec 0$). Poisson brackets will always be understood
to be taken with respect to the canonically conjugate variables $q_i$ 
and $p_i$. 
For the purpose of illustration we suppose that the second term on the right hand side vanishes on the surface defined by the primary constraints,
and that we have a new constraint $\phi^{(1)}_a = 0$, with
%%%%%%%%%%%%%%%%%%%%%%
\footnote{If this is not the case, the algorithm stops.}
%%%%%%%%%%%%%%%%%%%%%%% 
\be\label{constraint1}
\phi^{(1)}_a := \{\phi^{(0)}_a,H\}\,,
\ee
which is only a function of $q$ and $p$.
Hence from (\ref{constraint1}) and (\ref{vdotE1}),
\be 
\phi^{(1)}_{a} = \vec v^{(1)}(a)\cdot\vec E^{(1)}   
+\sum_{c} \lambda^{c} \{\phi^{(0)}_{c},\phi^{(0)}_{a}\} \,,
\ee
or
\be\label{Phi1}
\phi^{(1)}_{a} = \vec v^{(1)}(a)\cdot\vec E^{(1)} +  
\sum_{b,c} \lambda^{c}C^{[000]}_{cab}
(\vec v^{(0)}(b)\cdot\vec E^{(0)})\,,
\ee
where the structure functions $C^{[000]}_{cab}$ are defined by 
\be  
\{\phi^{(0)}_{c},\phi^{(0)}_{a}\} = 
\sum_{c}C^{[000]}_{cab}\phi^{(0)}_{b}\,,
\ee 
and use has been made of (\ref{vdotE0}). Note that $\phi^{(1)}_a$ is again a function of only $q$ and $p$.
  
We now repeat the process and adjoin the time derivative of 
the constraint (\ref{Phi1}) to the equations of motion to construct $\vec E^{(2)}$:
\be
{\vec E}^{(2)} = \left(
\begin{array}{c}
\vec E^{(0)}\\
\frac{d}{dt}\vec\phi^{(0)}\\
\frac{d}{dt}\vec\phi^{(1)}
\end{array}\right)\,,
\ee
where $\vec\phi^{(0)}$ and $\vec\phi^{(1)}$ are $N$  
component column vectors. This leads to a matrix ${\bf F}^{(2)}$. As we continue with this iterative process, the number of new zero modes generated at each new level will in general be reduced, as ``gauge identities" are being generated along the way
(see below). Hence the number of components  of $\vec\phi^{(\ell)}$ will
in general decrease, as the level $\ell$ increases.

The constraints $\phi^{(\ell)}_{a}$ with $\ell\ge 1$ can be iteratively 
constructed from the recursion relation, 
\be\label{identity}
\phi^{(\ell+1)}_{a} = 
\vec v^{(\ell+1)}(a)\cdot\vec E^{(\ell+1)} 
+\sum^{\ell}_{\ell'=0}\sum_{b,c}\lambda^{c}
C^{[0\ell\ell']}_{cab}
\phi^{(\ell')}_{b} \,, \quad  \ell \ge 0
\ee
where,
\be\label{iteration}
\phi^{(\ell+1)}_{a} := \{\phi^{(\ell)}_{a},H\}
\ee
and the sum over $b$ in (\ref{identity}) runs over all constraints $\phi^{(\ell')}_b$ at level $\ell'$,
%%%%%%%%%%%%%%
\footnote{Note that this set of constraints may be smaller in number than the set of primary constraints.}
%%%%%%%%%%%%%%
The coefficients $C^{[0\ell\ell']}_{cab}$ are structure functions defined by 
\be\label{structure-constants}  
\{\phi^{(0)}_{c},\phi^{(\ell)}_{a}\} = 
\sum^\ell_{\ell'=0}\sum_{b}C^{[0\ell\ell']}_{cab}
\phi^{(\ell')}_{b}\,.
\ee
The zero modes at level $\ell+1$ have the 
following generic form :
\be\label{generic-eigenvectors}
\vec v^{(\ell+1)}(b) = (-\tilde\nabla\phi^{(\ell)}_{b},\nabla\phi^{(\ell)}_{b},\vec 0,
\hat e^{(\ell)}(b)) \,,\quad  \ell \ge 0 \,.
\ee
Here  $\hat e^{(\ell)}(b)$ is a unit vector with the only non-vanishing component at the position of the 
constraint $\phi^{(\ell)}_{b}$ in the array $(\vec\phi^{(0)}, 
\vec\phi^{(1)},\cdot\cdot\cdot,\vec\phi^{(\ell)})$ appearing 
in the expression for $\vec E^{(\ell)}$.
The iterative process in a chain labelled by $``a"$ will come to a halt at level $\ell = N_a +1$, when
\be\label{final-element}
\phi^{(N_a+1)}_a = \{\phi^{(N_a)}_a,H\} = 
\sum_{\ell=0}^{N_a}\sum_{b}h^{[N_a\ell]}_{ab}\phi^{(\ell)}_{b}\,.
\ee
Making use of (\ref{identity}) and (\ref{final-element}), and setting $\ell = N_a$ in (\ref{identity}), this equation takes the form
\be\label{gaugeidentity}
G_a := \vec v^{(N_a+1)}(a)\cdot\vec E^{(N_a+1)} -\sum_{\ell=0}^{N_a}\sum_b K^{[N_a\ell]}_{ab}\phi^{(\ell)}_b \equiv 0\,,
\ee
where
\be\label{K-coefficient}
K^{[N_a\ell]}_{ab}= h^{[N_a\ell]}_{ab}-\sum_{c}\lambda^cC^{[0N_a\ell]}_{cab}\,.
\ee
Eq. (\ref{gaugeidentity}) expresses the fact, that 
the $\phi$-chains labelled by $``a^"$ ends in a ``gauge identity" at the level $N_a+1$.

Iteration of (\ref{identity}), starting with $\ell=0$, allows us to express all the constraints in terms of scalar products $\vec v^{(\ell)}\cdot\vec E^{(\ell)}$.
Substituting the resulting expressions into (\ref{gaugeidentity}), and  
multiplying each of the gauge identities $G_{a}\equiv 0$ by an arbitrary function of time $\alpha_{a}(t)$, 
the content of all the identities can be summarized by an 
equation of the form
\be\label{general-identity}
\sum_{a=1}^N\sum_{\ell=0}^{N_a+1}\rho^{(\ell)}_{a}(Q,\alpha)
\left(\vec v^{(\ell)}(a)\cdot\vec E^{(\ell)}\right) \equiv 0 \,,
\ee
where
\be\label{El}
{\bf \vec E}^{(\ell)} = \left(
\begin{array}{c}
\vec E^{(0)}\\
\frac{d}{dt}\vec\phi^{(0)}\\
\frac{d}{dt}\vec\phi^{(1)}\\
\cdot\\
\cdot\\
\cdot\\
\frac{d}{dt}\vec\phi^{(\ell-1)}
\end{array}\right)\,.
\ee

Now, because of the generic structure of the eigenvectors (\ref{generic-eigenvectors}) we have from (\ref{El})
\be\label{vdotEl}
\vec v^{(\ell+1)}(a)\cdot \vec E^{(\ell+1)} = \sum^{2n}_{\alpha=1}
\left(v^{(\ell+1)}_\alpha(a) E^{(0)}_\alpha\right) 
+\frac{d\phi^{(\ell)}_{a}}{dt}
\,,\quad \ell = 0,\cdots,N_a
\ee
where $n$ is the number of coordinate degrees of freedom, and the constraints 
$\phi^{(\ell)}_a$ appearing on the RHS can be expressed, by iterating (\ref{identity}), in terms 
of scalar products $\vec v^{(k)}\cdot\vec E^{(k)}$, which in turn can be decomposed in the form (\ref{vdotEl}). Upon making 
a sufficient number of ``partial decompositions" $udv = d(uv) -vdu$, the identity (\ref{general-identity}) can be written in the form
\be\label{gaugeidentity2}
\sum_{a=1}^{N}\sum^{N_a}_{\ell=0}\epsilon^{(\ell)}_{a}\sum_{\alpha=1}^{2n+N}
\left(v^{(\ell+1)}_\alpha(a) E^{(0)}_\alpha\right)
+ \sum_{a=1}^N \tilde\epsilon_a \sum^{2n+N}_{\alpha=1}(v^{(0)}_\alpha(a)E^{(0)}_{\alpha})-\frac{dF}{dt}
\equiv 0\,,
\ee
where the 
$\{\epsilon^{(\ell)}_{a}\}$ and $\tilde\epsilon_a$ depend on the $N$ 
arbitrary functions of time $\{\alpha_{a}(t)\}$, as well as on the $Q_\alpha$'s 
and time derivatives thereof.  
The expression is of the form
%%%%%%%%%%%%%%%
\footnote{The minus sign has been introduced to cast the transformation 
laws in a standard form, when written in terms of Poisson brackets.}
%%%%%%%%%%%%%%%%
\be\label{EdeltaQ}
-\sum_\alpha E^{(0)}_\alpha\delta Q_\alpha = \frac{dF}{dt}\,,
\ee
where
\be\label{deltaQ}
\delta Q_\alpha = -\sum^N_{a=1}\sum^{N_a}_{\ell=0} {\cal\epsilon}^{(\ell)}_{a}
v^{(\ell+1)}_\alpha (a)
- \sum_{a=1}^N \tilde\epsilon_a v^{(0)}_\alpha(a) \,.
\ee
For infinitessimal  ${\cal\epsilon}^{(\ell)}_{a}$ the time integral of the LHS of (\ref{EdeltaQ}) is just the variation of the total action (\ref{delta-Stotal}).
Hence we conclude that the transformations (\ref{deltaQ}) 
leave the total action (\ref{Stotal}) invariant.
But because of the generic structure of the eigenvectors (\ref{generic-eigenvectors}) we have  from (\ref{deltaQ}),
\be\label{deltaq}
\delta q_i = \delta Q_i = \sum^N_{a=1}\sum^{N_a}_{\ell=0}{\cal\epsilon}^{(\ell)}_{a}
\frac{\partial\phi^{(\ell)}_{a}}{\partial p_i} 
= \sum^N_{a=1}\sum_{\ell=0}^{N_a}{\cal\epsilon}^{(\ell)}_{a}
\{q_i,\phi^{(\ell)}_{a}\}\,\,,i=1,\cdots,n\,.
\ee
and
\be\label{deltap}
\delta p_i = \delta Q_{n+i} 
= - \sum^N_{a=1}\sum^{N_a}_{\ell=0}{\cal\epsilon}^{(\ell)}_{a}
\frac{\partial\phi^{(\ell)}_{a}}{\partial q_i} 
=\sum^N_{a=1}\sum_{\ell=0}^{N_a}{\cal\epsilon}^{(\ell)}_{a}
\{p_i,\phi^{(\ell)}_{a}\}\,,
\ee
while
\be\label{deltalambda1}
\delta\lambda^{a} = \delta Q_{2n+a} = -\tilde\epsilon_a \,.
\ee
 Note that according to (\ref{iteration}) the constraint $\phi^{(\ell+1)}$ has been generated from  $\phi^{(\ell)}$ via the Poisson bracket of  $\phi^{(\ell)}$ with $H$. Hence, given the 
primary constraints, the form of the secondary constraints are fixed,
and one should not modify their structure, replacing them by an equivalent set of constraints.
%%%%%
\footnote{Thus, for example, one should not replace a constraint $p^2 = 0$ by $p=0$.}
%%%%%
We have thus proven a proposition made in \cite{Stefano} regarding the form of the 
first class constraints that are generators of local symmetries.
 
Finally we would like to emphasize, that we have implicitely asumed throughout that the structure functions appearing in equations 
(\ref{structure-constants}) and (\ref{final-element}) are finite on the constraint surface. If this is not the case, we expect that our algorithm 
will still generate the correct local {\it off-shell} symmetries of the total action, if 
the gauge identities are non-singular on the constrained surface, while 
they do not implement symmetries on the level of the Hamilton equations of 
motion. This will be demonstrated in our example 2 in the section 4.  
In the following section we first illustrate our formalism for the general case of two primary constraints and one secondary constraint.

%%%%%%%%%%%%%%%%%%%%%%%%%%%%%%%%%%%%%%%%%%%%%%%%%%%%%%%
\section{General system with two primaries and one secondary constraint}
%%%%%%%%%%%%%%%%%%%%%%%%%%%%%%%%%%%%%%%%%%%%%%%%%%%%%%%

In this section we apply the formalism to the case where the system 
exhibits two primary constraints $\phi^{(0)}_{a}, a = 1,2$, and one secondary constraint $\phi^{(1)}_2$, which in Dirac's language is generated 
from the presistency in time of $\phi^{(0)}_2$. (This is illustrated by the table below.) The specific form of 
the Hamiltonian is irrelevant. {\it The constraints are assumed to be first class.}
\be
\begin{array}{c|c}
\phi^{(0)}_1&\phi^{(0)}_2\\ 
\hline
&\phi^{(1)}_2
\nonumber
\end{array}
\ee
 
The total Lagrangean $L_T$ is given by (\ref{Ltotal}) with $N=2$. Proceeding as in section 2, the primary 
constraints can be written in the form (\ref{vdotE0}), where $\vec v^{(0)}(1) 
= (\vec 0,\vec 0,1,0)$ and  $\vec v^{(0)}(2) = (\vec 0,\vec 0,0,1)$. 

Next we 
construct the vector (\ref{E1}), i.e. 
$\vec E^{(1)} = (\vec E^{(0)},\dot\phi^{(0)}_1,\dot\phi^{(0)}_2)$, and 
 the zero modes of the corresponding matrix ${\bf F}^{(1)}$, which are given by (\ref{zeromode1}), i.e., $\vec v^{(1)}(1) = (-\tilde\nabla\phi^{(0)}_1,\nabla\phi^{(0)}_1,0,0,1,0)$, 
$\vec v^{(1)}(2) = (-\tilde\nabla\phi^{(0)}_2,\nabla\phi^{(0)}_2,0,0,0,1)$.     Since 
only $\phi^{(0)}_2$ is assumed to lead to a new constraint, we are immediately left with one gauge identity at level 1, generated from $\phi^{(0)}_1$:
\be
G_1=\vec v^{(1)}(1)\cdot\vec E^{(1)} -\sum_{b=1}^2K^{[00]}_{1b}\phi^{(0)}_b
\equiv 0 \,,
\nonumber\\
\ee
or 
\be\label{GI1}
G_1=\vec v^{(1)}(1)\cdot\vec E^{(1)} -\sum_{b=1}^2K^{[00]}_{1b}\left(\vec v^{(0)}(b)\cdot\vec E^{(0)}\right) \equiv 0\,,
\ee   
which is nothing but (\ref{gaugeidentity}) with $N_1=0$. On the other hand $\phi^{(0)}_2$ gives rise, by assumption, to a new constraint $\phi^{(1)}_2$ at level 1, which is given by 
\be
\phi^{(1)}_2 = \vec v^{(1)}(2)\cdot\vec E^{(1)} + 
\sum^2_{b,c=1}\lambda^{c}C^{[000]}_{c2b}
\left(\vec v^{(0)}(b)\cdot\vec E^{(0)}\right)\,.
\ee
We are therefore led to construct 
\be
{\vec E}^{(2)} = \left(
\begin{array}{c}
\vec E^{(0)}\\
\frac{d}{dt}\phi^{(0)}_1\\
\frac{d}{dt}\phi^{(0)}_2\\
\frac{d}{dt}\phi^{(1)}_2
\end{array}\right)
\ee
as well as the corresponding rectangular matrix $F^{(2)}$ and its left zero modes, of 
which only the contraction of 
\be
\vec v^{(2)}(2) = (-\tilde\nabla\phi^{(1)}_2,\nabla\phi^{(1)}_2,0,0,0,0,1)
\ee
with $\vec E^{(2)}$ leads to a new equation, which is necessarily a 
gauge identity, since we have assumed that the system only possesses one 
secondary constraint $\phi^{(1)}_2$. The gauge identity at level 2 has the form
\be
G_2=\vec v^{(2)}(2)\cdot\vec E^{(2)}-\sum_{b=1}^2
K^{[10]}_{2b}\phi^{(0)}_{b}
- K^{[11]}_{22}\phi^{(1)}_2 \equiv 0 \,,\nonumber
\ee
or
\ba\label{GI2}
G_2=\vec v^{(2)}(2)\cdot\vec E^{(2)}&-&\sum^2_{b=1}K^{[10]}_{2b}
\left(\vec v^{(0)}({b})\cdot\vec E^{(0)}\right)\\
&-& K^{[11]}_{22}\left(\vec v^{(1)}(2)\cdot\vec E^{(1)}
+\sum_{b,c=1}^2\lambda^{c}C^{[000]}_{c2b}\left(\vec v^{(0)}(b)\cdot\vec E^{(0)}\right)\right) \equiv 0\,.\nonumber
\ea
Multiplying the gauge identities (\ref{GI1}) and (\ref{GI2}) by the arbitrary  
functions $\alpha_1(t)$ and $\alpha_2(t)$, respectively, and taking their sum, the resulting expression can be written in the form (\ref{general-identity}). Upon making a sufficient 
number of ``partial differential decompositions", one finds, after some
algebra, that the information encoded in 
the gauge identities can be written in the form (\ref{gaugeidentity2}), 
where
\ba
\epsilon^{(0)}_1 &=& -\alpha_1\cr
\epsilon^{(1)}_2 &=& -\alpha_2\cr
\epsilon^{(0)}_2 &=& \dot\alpha_2 +\alpha_2K^{[11]}_{22}\cr
\delta\lambda^a &=&  
-\dot\alpha_2\sum_{c}\lambda^{c}C^{[000]}_{c2a} - 
\alpha_2K^{[10]}_{2a}
- \alpha_1 K^{[00]}_{1a}\cr
 &-& \alpha_2K^{[11]}_{22}\sum_{c}\lambda^{c}C^{[000]}_{c2a}
+\ddot\alpha_2\delta_{2a} - \dot\alpha_1\delta_{1a} 
+\frac{d}{dt}\left(\alpha_2K^{[11]}_{22}\right)\delta_{2a}\,.
\ea
are the only non-vanishing parameters.
The corresponding transformation laws for the coordinates $q_i$ and 
momenta $p_i$ 
are given by (\ref{deltaq}) and (\ref{deltap}).

Having obtained $\epsilon^{(0)}_{1},\epsilon^{(0)}_{2},
\epsilon^{(1)}_{2}$ and $\delta\lambda^a$ expressed 
in terms of $\alpha_1(t)$ and $\alpha_2(t)$, one now verifies 
that the $\epsilon^{(\ell)}_a$  are solutions to the well 
known recursion relations following from the requirement that the transformation laws (\ref{deltaq}) and (\ref{deltap}) be symmetries of the total action\cite{Girotti,Henneaux,BRR}:
\be\label{recursion-relations}
{\dot\epsilon}^{(\ell)}_{a}  
+\sum_{\ell'}\sum_{b}{\cal\epsilon}^{(\ell')}_{b}K^{[\ell'\ell]}_{ba} = 0 
\,,\quad \ell\ge 1
\ee
and
\be\label{deltalambda}
\delta\lambda^{a} = {\dot{\cal\epsilon}}^{(0)}_{a} 
+ \sum_{\ell'}\sum_{b}{\cal\epsilon}^{(\ell')}_{b}K^{[\ell'0]}_{ba}\,.
\ee
In our case these equations reduce to 
\be\label{Ex-recursion1}
\frac{d\epsilon^{(1)}_2}{dt} + \sum^2_{b=1}\epsilon^{(0)}_{b}
K^{[01]}_{b2}+\epsilon^{(1)}_2 K^{[11]}_{22} = 0\,,
\ee
and
\be\label{Ex-recursion2}
\delta\lambda^a = \dot\epsilon^{(0)}_a 
+ \sum_b \epsilon^{(0)}_b K^{[00]}_{ba} + \epsilon^{(1)}_2 K^{[10]}_{2a}\,,
\ee
where we have made use of the fact that, because of the way in which the constraints
have been generated, $K^{[\ell1]}_{a1} = 0\ (a=1,2)$, $K^{[01]}_{22} = 1$,
$K^{[01]}_{12} = 0$, $C^{[001]}_{abc} = 0$, and
$K^{[00]}_{22} = - \sum_c \lambda^c C^{[000]}_{c22}$. One then verifies that the above equations are indeed satisfied.

%%%%%%%%%%%%%%%%%%%%%%%%%%%%%%%%%%%%%%%%%%%%%%%%%%%% 
\section{Counterexamples to Dirac's conjecture?}
%%%%%%%%%%%%%%%%%%%%%%%%%%%%%%%%%%%%%%%%%%%%%%%%%%%%

In the literature it has been stated that Dirac's conjecture is not always
correct \cite{Cawley,Teitelboim,Zanelli}.
In the following we present three models which have served as examples for this 
in the literature and show, that there is no clash with Dirac's conjecture if the first class constraints are understood to be generated in the definite iterative sense discussed in section 2.

\bigskip\noindent
%%%%%%%%%%%%%%%%%%%%%%%%%%%%%%%%%%%%%%%
{\bf Example 1}
%%%%%%%%%%%%%%%%%%%%%%%%%%%%%%%%%%%5%%%

\bigskip 
An example considered in \cite{Teitelboim} is given by the Lagrangean
\be\label{conjecture-Lagrangean}
L = \frac{1}{2}e^{q_2}{\dot q_1}^2\,.
\ee
In the following we analyze this system in detail, and point out some 
subtleties leading to an apparent clash with Dirac's conjecture.

The Lagrange equations of motion can be summarized by a single equation
\be\label{qdot}
\dot q_1 = 0\,.
\ee
Hence $q_2$ is an arbitrary function. 

Eq. (\ref{qdot}) does not possess a local symmetry. On the Hamiltonian level the system nevertheless 
exhibits two first class constraints, which, as we now show induce 
transformations which are off-shell symmetries 
of the total action $\int dt\ L_T$, with $L_T$ defined in defined in (\ref{Ltotal}). However only one of the constraints generates a symmetry of the Hamilton equations of motion.  

From (\ref{conjecture-Lagrangean}) we obtain for the (only) primary constraint
\be\label{conjecture-primary}
\phi^{(0)} = p_2 = 0\,.
\ee
The canonical Hamiltonian evaluated on the primary surface is given by
\be
H(q,p) = \frac{1}{2}e^{-q_2}p_1^2\,,
\ee
and the first order total Lagrangean reads
\be\label{Lex2}
L_T(q,p,\lambda;\dot q,\dot p,\dot\lambda) =  \sum_i p_i\dot q_i-H(q,p)
-\lambda \phi^{(0)}\,.
\ee
Considered as a function of the "coordinates" $q_i$, $p_i$ and $\lambda$ and 
their time derivatives, it has the form (\ref{Ltotal-general}), where
$2n+N=5$, $Q_\alpha = (\vec q, \vec p,\lambda)$, 
and  $a_i = p_i$ ($i=1,2$) for the non-vanishing components of
$a_\alpha$. 

The Euler derivatives are given by (\ref{Euler-derivative}), where
$F^{(0)}_{\alpha\beta}=\partial_\alpha a_\beta - \partial_\beta a_\alpha$  
are the elements of a $5\times 5$ matrix with non-vanishing components
$F_{13} = F_{24} = -F_{31} = -F_{42} = -1$. 
The equations of motion are given by $E^{(0)}_\alpha = 0$, and 
yield the Hamilton equations of motion
\ba\label{Hamilton-equations}
\dot q_1 &-& e^{-q_2}p_1 = 0\cr
\dot q_2 &-& \lambda = 0\cr
\dot p_1 &=& 0\cr
\dot p_2 &-& \frac{1}{2}e^{-q_2}p_1^2 = 0\,,
\ea
as well as the constraint (\ref{conjecture-primary}).
We now proceed with the construction of the constraints and gauge identities 
as described in section 2. Since we have only one primary constraint, the 
formalism simplifies considerably. 

The matrix $F^{(0)}_{\alpha\beta}$ has one left zero-mode $\vec v^{(0)} = (0,0,0,0,1)$, whose contration with 
$\vec E^{(0)}$ just reproduces on shell the primary constraint:
$\vec v^{(0)}\cdot\vec E^{(0)} = \phi^{(0)}$. 
Proceeding in the manner described in section 2, we construct $\vec E^{(1)}$ and corresponding left eigenvector of $F^{(1)}$, $v^{(1)} = (0,-1,0,0,0,1)$,
leading to the secondary constraint
$\vec v^{(1)}\cdot\vec E^{(1)} = \phi^{(1)}$,
where
\be\label{E1ex}
\vec E^{(1)} = \left(
\begin{array}{r}
\vec E^{(0)}\\
\frac{d\phi^{(0)}}{dt}\\
\end{array}\right)
= \left(
\begin{array}{c}
\vec E^{(0)}\\
\frac{d}{dt}(\vec v^{(0)}\cdot\vec E^{(0)})\\
\end{array}\right)\,,
\ee
and
\be
\phi^{(1)} = \frac{1}{2}e^{-q_2}p^2_1 \,.
\ee
Note that by construction
$\phi^{(1)} = 0$, on\ shell.
Since $\{\phi^{(0)},\phi^{(1)}\} = \phi^{(1)}$, $\phi^{(0)}$ and
$\phi^{(1)}$ form a first class system.
The algorithm is found to stop at level ``2" where the following gauge identity 
is generated:
\be
\vec v^{(2)}\cdot\vec E^{(2)} + 
\lambda \vec v^{(1)}\cdot\vec E^{(1)} \equiv 0\,,
\ee
with
\be
\vec v^{(2)} = (-e^{-q_2}p_1,0,0,-\frac{1}{2}e^{-q_2}p^2_1,0,0,1)\,,
\ee
and  
\be\label{E2ex}
\vec E^{(2)} = \left(
\begin{array}{r}
\vec E^{(0)}\\
\frac{d\phi^{(0)}}{dt}\\
\frac{d\phi^{(1)}}{dt}
\end{array}\right)
= \left(
\begin{array}{c}
\vec E^{(0)}\\
\frac{d}{dt}(\vec v^{(0)}\cdot\vec E^{(0)})\\
\frac{d}{dt}(\vec v^{(1)}\cdot\vec E^{(1)})
\end{array}\right)\,.
\ee
The gauge identity can be reduced to the form 
\be
v^{(2)}_\alpha E^{(0)}_\alpha + \frac{d}{dt}(v^{(1)}_\alpha E^{(0)}_\alpha) 
+\frac{d^2}{dt^2}(v^{(0)}_\alpha E^{(0)}_\alpha) 
+\lambda[(v^{(1)}_\alpha E^{(0)}_\alpha) + 
\frac{d}{dt}(v^{(0)}_\alpha E^{(0)}_\alpha)] \equiv 0\,,
\ee
where a summation over $\alpha = 1,\cdot\cdot\cdot,5$ is understood.
Multiplying this expression by an arbitrary function of time $\epsilon(t)$, 
this identity becomes of the form (\ref{EdeltaQ}), with 
\be\label{deltaQalpha}
\delta Q_\alpha = \epsilon v^{(2)}_\alpha + (\lambda\epsilon-\dot\epsilon)v^{(1)}_\alpha + 
\left(\ddot \epsilon - \frac{d}{dt}(\lambda\epsilon)\right)v^{(0)}_\alpha\,.
\ee
In terms of the Hamiltonian variables, $q_i$, $p_i$ and $\lambda_i$, (\ref{deltaQalpha}) 
implies the following transformation laws
\ba\label{conjecture-deltaQ}
\delta q_1 &=& -\epsilon e^{-q_2}p_1 = \epsilon^{(1)}\{q_1,\phi^{(1)}\}\,,\cr
\delta q_2 &=& = \dot\epsilon - \lambda\epsilon =  \epsilon^{(0)}\{q_2,\phi^{(0)}\}\,,\cr
\delta p_1 &=& 0\,,\cr
\delta p_2 &=& -\frac{1}{2}\epsilon e^{-q_2}p_1^2 = \epsilon^{(1)}\{p_2,\phi^{(1)}\}\,,\cr
\delta \lambda &=& \ddot\epsilon - \frac{d}{dt}(\lambda\epsilon) 
= \dot\epsilon^{(0)}\,,
\ea
where
\be
\epsilon^{(0)} =\dot\epsilon - \lambda\epsilon\,,\quad
\epsilon^{(1)} = -\epsilon \,.
\ee
One readily verifies, that the $\epsilon^{(\ell)}$ satisfy recursion relations (\ref{recursion-relations}).

We have thus verified that the transformations correspond to a symmetry of 
the action (\ref{Stotal}). This symmetry is realized {\it off shell}, 
and requires the 
full set of transformation laws (\ref{conjecture-deltaQ}). On the other hand, 
the first class constraint ${\tilde\phi}^{(1)}:=p_1 = 0$, although equivalent to the secondary constraint ${\phi}^{(1)}=0$, is not a generator of a local symmetry
of the total action. Indeed, as observed in \cite{Teitelboim},  $p_1$ induces translations in $q_1$, which is not a symmetry of the total Lagrangean (\ref{Lex2}). 
Note further that on the constrained surface,  the variations 
of $\delta q_1$ and 
$\delta p_2$ vanish, and hence do not generate a symmetry on the level of the Hamilton equations of motion. Thus $\phi^{(1)}$ becomes and ineffective generator on this level. The remaining symmetry is just the statement
$\delta\dot q_2 = \delta\lambda$, which is consistent with (\ref{Hamilton-equations}).
It has no analogue on the level of the Lagrange equations of motion
following from (\ref{conjecture-Lagrangean}).

\bigskip\noindent
%%%%%%%%%%%%%%%%%%%%%%%%%%%%%%%%%%%%%%%%%%%%%%%%%%%%%%%%%
{\bf Example 2: constraint of the form $\phi = f^k = 0$.}
%%%%%%%%%%%%%%%%%%%%%%%%%%%%%%%%%%%%%%%%%%%%%%%%%%%%%%%%%

\bigskip
The following example taken from ref. \cite{Zanelli} illustrates our comment made at the end of section 2
regarding the case where the structure functions (\ref{K-coefficient})
are singular on the constrained surface, while the
gauge identitites are well defined on that surface.

Consider the Lagrangean,
\be\label{LZanelli}
L = \frac{1}{2}\dot q^2 + uf^k(q)\,,\quad k > 1\,,
\ee
which has been classified as being of ``type II" in ref. \cite{Zanelli}.
%%%%%%%%%%
\footnote{In order to simplify the discussion we have restricted ourselves
to two degrees of freedom, $q$ and $u$.}
%%^%%%%%%%
The canonical Hamiltonian evaluated on the primary surface $p_u=0$ is given by 
$H=\frac{1}{2}p^2 - uf^k(q)$, and correspondingly the
first order Lagrangean reads
\be\label{LTZanelli}
L_T(Q,\dot Q) = p\dot q + p_u \dot u - \frac{1}{2}p^2 + uf^k(q) - \lambda p_u\,,
\ee
where $Q = (q,u,p,p_u,\lambda)$. The Euler-Lagrange equations derived from 
(\ref{LTZanelli}) are just the Hamilton equations of motion (including the
primary constraint $p_u=0$).
Following the above Lagrangean algorithm, we obtain at levels $\ell = 0,1,2$,
\ba
v^{(0)} &=& (0,0,0,0,1)\,,\quad \phi^{(0)} = \vec v^{(0)}\cdot \vec E^{(0)}=p_u\,,\label{v0}\\
v^{(1)} &=& (0,-1,0,0,0,1)\,,\quad \phi^{(1)} = \vec v^{(1)}\cdot \vec E^{(1)}=f^k(q)\,,\label{v1}\\
v^{(2)} &=& (0,0,f^{k-1}f',0,0,0,1)\,,\quad\vec v^{(2)}\cdot \vec E^{(2)}
=kp(\ln f)'\phi^{(1)}\,,\label{v2}
\ea
where the prime denotes the derivative with respect to $q$. Note that for 
$k > 1$ the right hand side of (\ref{v2}) is well defined on the constrained surface $\phi^{(1)}=0$, and in fact vanishes there.
Hence at level $\ell=2$ we have generated a gauge identity, which upon making use of $\phi^{(1)} = \vec v^{(1)}\cdot \vec E^{(1)}$ takes the form,
\be\label{identity-ex2}
\vec v^{(2)}\cdot \vec E^{(2)} - kp(\ln f)'\vec v^{(1)}\cdot \vec E^{(1)}=0\,.
\ee
In the present example the structure functions in (\ref{structure-constants})
vanish identically, and the ``coordinate" dependent factor $kp(\ln f)'$ in (\ref{identity-ex2}) is  nothing but the structure
function $K$ in (\ref{K-coefficient}). This function is singular on the constrained surface. Nevertheless the gauge identity (\ref{identity-ex2}) ``generates" {\it off-shell} the correct local symmetry of the total action, as we
demonstrate below.
 
In the present example $\vec E^{(1)}$ and  $\vec E^{(2)}$ have again the 
form of (\ref{E1ex}) and (\ref{E2ex}), respectively. 
%Hence 
%\ba
%\vec v^{(1)}\cdot E^{(1)} &=& \sum_{\alpha=1}^5 v^{(2)}_\alpha E^{(0)}_\alpha
%+ \frac{d}{dt}(\vec v^{(0)}\cdot E^{(0)})\,.\\
%\vec v^{(2)}\cdot \vec E^{(2)} &=& \sum_{\alpha=1}^5 v^{(2)}_\alpha E^{(0)}_\alpha
%+\frac{d}{dt}(\vec v^{(1)}\cdot \vec E^{(1)})\,.
%\ea
Making use of these relations, and multiplying the gauge identity
(\ref{identity-ex2}) by $\epsilon$, one finds that the resulting identity can be written in the form 
(\ref{EdeltaQ}) with $\delta Q_{\alpha}$ the infinitesimal transformations,
\ba\label{Z-inftransf}
\delta q &=& 0\,,\\
\delta u &=& \dot\epsilon + \epsilon kp(\ln f)'\,,\nonumber\\
\delta p &=& \epsilon k f^{k-1}f'\,,\nonumber\\
\delta p_u &=& 0\,,\\
\delta \lambda &=& \ddot\epsilon + \frac{d}{dt}(\epsilon k p(\ln f)')\,.\nonumber
\ea
These transformations are only defined off-shell.
Computing the corresponding variation $\delta L_T$ one finds
that it is given by a total time derivative 
%%%%%%%%
\footnote {For the second order Lagrangean the corresponding gauge
identities are found to imply the infinitesimal symmetry transformation,
\be
\delta q = 0\,,\quad 
\delta u = \dot\epsilon + \epsilon kp(\ln f)'\,.\nonumber
\ee
One again verifies in this case that $\delta L = 
\frac{d}{dt}(\epsilon f^k(q))\,.$}
%%%%%%%%
:
\be
\delta L_T = \frac{d}{dt}(\epsilon f^k(q))\,.
\ee
In accordance with our expectations, one verifies that $\delta Q_\alpha$ 
can be written in the form
\be
\delta Q_\alpha = \sum_{\ell=0}^1\epsilon_\ell\{Q_\alpha,\phi^{(\ell)}\}\,, \quad
\alpha =1,\cdots,4\,,
\ee
where
\ba
\epsilon_{0} &=& \dot\epsilon + \epsilon kp(\ln f)'\,,\\
\epsilon_{1} &=& -\epsilon\,.
\ea
For the Lagrange multiplier one has $\delta \lambda = \frac{d}{dt}\epsilon_0$. 
It is then an easy matter to verify that $\epsilon_\ell$ and $\delta\lambda$ 
indeed satisfy equations (\ref{recursion-relations}) and (\ref{Ex-recursion2}).

The above infinitesimal transformations (\ref{Z-inftransf})
represent symmetry transformations of the total action, away
from the constraint surface $f^k(q)=0$. They are not defined
on the surface $\phi^{(1)}=0$.  Correspondingly they do not 
represent symmetries of the Hamilton equations of motion. 

Summarizing, the above example shows that if we generate the constraints
in a strict iterative way following the Lagrangean algorithm, then the first
class constraints do generate the local {\it off shell} symmetries of the total action
in the form conjectured by Dirac, if the gauge identities are finite 
expressions on the constraint surface, although the structure functions 
may be singular there. We have, however, no proof that this is generally true.

%Notice that in the above considerations the constraints $\phi^{(\ell)}$ 
%have been generated in a strict iterative way. 
%In particular, we did not replace the constraint $\phi^{(1)} = f^k = 0$, %$(k>1)$, by its equivalent expression $f = 0$. Had we replaced at level %$\ell=1$ the constraint $f^k(q)=0$ by $f(q)=0$, the algorithm would not stop at %this point, but rather reduce the system to a mixed system of one first class %constraint $p_u = 0$, and an infinite number of second class constraints.
%The remaining first class constraint however no longer generates a symmetry
%of the Lagrangean $L$.
%lead in general to a further constraint, reducing the system to a
%second class system, lacking any gauge symmetry \cite{Zanelli}, contrary to our 
\bigskip\noindent
%%%%%%%%%%%%%%%%%%%%%
{\bf Example 3: bifurcations}
%%%%%%%%%%%%%%%%%%%%%

\bigskip
In the following we now consider an example of a system 
exhibiting bifurcations of constraints, such as considered in ref. \cite{Lusanna}, and show that also in this case our algorithm generates 
correctly the gauge symmetries of the total 
action, whereas the first class constraints corresponding to the choice of a
particular branch of these bifurcations are not part of our iterative algorithm, and do not generate gauge symmetries of the 
total action.

Consider the so called ``Christ-Lee model" \cite{Christ-Lee} discussed in ref. \cite{Lusanna}, and defined by the Lagrangean
\be
L = \half\dot {q}_1^2 + \half \dot{q}_2^2 - q_3(q_1\dot{q}_2 - q_2\dot{q}_1)
- V(q_1^2 + q_2^2) + \frac{1}{2}q_3^2(q^2_1+q^2_2)\,,
\ee
with $V(x)$ some potential. 
The equivalent first order Lagrangean in our approach reads,
\be
L_T = \sum_i p_i\dot q_i - \half p_1^2 - \half p_2^2 - q_3(q_1p_2 - q_2p_1) 
- V - \lambda p_3\,.
\ee
The corresponding Euler-Lagrange equations take the form (\ref{Euler-derivative})
%\be
%E^{(0)}_\alpha := -\sum_\alpha F^{(0)}_{\alpha,\beta} \dot Q_\beta %+K^{(0)}_\alpha
%\ee
where $Q_\alpha : (q_1,q_2,q_3,p_1,p_2,p_3,\lambda)$ 
and the $E_\alpha^{(0)}$'s are given by
\bear
E_1^{(0)} &=& \dot p_1 + p_2q_3 + \partial_1 V \,,\quad
E_2^{(0)} = \dot p_2 - p_1q_3 + \partial_2 V \,,\\
E_3^{(0)} &=& \dot p_3 + (q_1p_2 - q_2p_1)\,,\quad
E_4^{(0)} = -\dot q_1 + p_1 - q_2q_3\,,\\
E_5^{(0)} &=& -\dot q_2 + p_2 + q_1q_3\,,\quad 
E_6^{(0)} = -\dot q_3 + \lambda\,,\quad 
E_7^{(0)} =  p_3\,.
\eear
Going through the procedure of section 2 we arrive at a level zero and level one
constraint,
\be
\phi^{(0)} := \vec v^{(0)}\cdot \vec E^{(0)} = p_3 = 0\,,\quad 
\phi^{(1)} := \vec v^{(1)}\cdot \vec E^{(1)} = q_2p_1 - q_1p_2 = 0\,,
\ee
where $\vec v^{(0)} = (0,0,0,0,0,0,1)$ and 
$\vec v^{(1)} = (0,0,-1,0,0,0,0,1)$, and $\vec E^{(1)}$ has the form
(\ref{E1ex}).  
The iterative process stops at level two, with the gauge identity
$\vec v^{(2)}\cdot \vec E^{(2)} \equiv 0$. ${\vec E}^{(2)}$ has the form (\ref{E2ex}) with
$v^{(2)} = (-q_2,q_1,0,-p_2,p_1,0,0,0,1)$ the level 2 zero mode. The simple form of this identity reflects
the fact that the structure functions $K^{[2\ell]}_{ab}$ appearing in 
(\ref{K-coefficient}) vanish for the case in question. Explicitely
one has
\be\label{Ex2-Gaugeidentity}
\vec v^{(2)}\cdot \vec E^{(2)}=v^{(2)}_\alpha{E}_\alpha^{(0)} 
+ \frac{d}{dt}v^{(1)}_\alpha{E}_\alpha^{(0)} 
+ \frac{d^2}{dt^2}v^{(0)}_\alpha{E}_\alpha^{(0)} \equiv 0.
\ee
This equation, when multiplied by $\epsilon(t)$, can be written in the form (\ref{EdeltaQ}) with
$F = \dot\epsilon \phi^{(0)} - \epsilon\phi^{(1)}$.
The $\{\delta Q_\alpha\}$ can be compactly written as
\be
\delta q_\alpha = \sum_{\ell=0}^1 \epsilon_\ell\{q_\alpha,\phi^{(\ell)}\} \,,\quad \delta \lambda = \ddot\epsilon\,,
\ee
where $\epsilon_0 = \dot\epsilon$, $\epsilon_1 = -\epsilon$.  

The constraints $\phi^{(0)} = 0$ and  $\phi^{(1)} = 0$
are first class and are thus found to generate an off-shell gauge symmetry of the total action,  in agreement with Dirac's conjecture.

Suppose now that we had chosen instead of $\phi^{(1)} = 0$ any one of the 
bifurcations: i) $q_1 = q_2 = 0$, ii)  $p_1 = p_2 = 0$,
iii) $q_1 = p_1 = 0$, and iv) $q_2 = p_2 = 0$. Consider for instance the case i) %%%%%
\footnote{ We are leaving herewith our algorithm, since with the choice of a particular branch the constraint can no longer be written in the form $\vec v^{(1)}\cdot\vec E^{(1)}$.}
%%%%%
.
Then we continue at level one with the constraints,
$\bar\phi^{(1)}_1 := q_1 = 0\,\quad \bar\phi^{(1)}_2 := q_2 = 0$.
The requirement of persistence in time then leads to two new constraints at level two:
$\bar\phi^{(2)}_1 = p_1 = 0\,,\quad \bar\phi^{(2)}_2 := p_2 = 0$.
Since $\dot{\bar\phi}_1^{(2)} \approx 0$ and $\dot{\bar\phi}_2^{(2)} \approx 0$
there are no further constraints.
We thus see that the system of constraints has been reduced to a mixed system with
two second class constraints, and just one first class constraint
$\phi^{(0)} := p_3 = 0$. Correspondingly we are led to consider 
$\bar G = \epsilon p_3$ as 
a potential generator for a local symmetry. One readily verifies that all variations vanish except for $\delta q_3$ and $\delta\lambda$. Hence
\be
\delta L_T = (\dot\epsilon - \delta\lambda)p_3 - \epsilon (q_1p_2 - q_2p_1)
\ee
To have asymmetry we must demand that
$\delta\lambda = \dot\epsilon$ and $q_1p_2 - q_2p_1 = 0$,
so that $L_T$ only exhibits a local invariance on the restricted surface
defined by the {\it first class} constraint generated at level 1 by the algorithm of section 2.  Hence $\bar G$ does not generate an off-shell symmetry of the total action. The Hamilton equations of motion, however, exhibit a symmetry $q_3 \to q_3 + \epsilon$, $\lambda \to \lambda + \dot\epsilon$.

%%%%%%%%%%%%%%%%%%%%%
\section{Conclusion}
%%%%%%%%%%%%%%%%%%%%%
In this paper we have first generalized the work of ref. \cite{Heinz} 
to arbitrary  first class Hamiltonian systems. Using purely Lagrangean methods we have derived the local symmetries of the total action, 
and have discussed in which sense the Dirac conjecture holds. More precisely, 
the Dirac conjecture was found to hold for first class constraints that have been generated in a definite, strict iterative way by the Lagrangean algorithm described in section 2. This provides a proof of the prescription given in ref. \cite{Stefano} regarding the form of the first class secondary constraints generating a local symmetry. 
In particular, we have quite generally shown  that the infinitesimal gauge transformations which leave
the total action invariant
have the form proposed by Dirac, i.e., 
$\delta q_i = \sum_A\epsilon^A\{q_i,\phi_A\}$, 
$\delta p_i = \sum_A\epsilon^A\{p_i,\phi_A\}$, where the $\{\phi_A\}'s$ are all the first class constraints generated by the algorithm refered to above. 
On the other hand, the replacement of constraints generated iteratively by an equivalent set of 
constraints is, in general, not allowed, and in fact may obliterate the 
full symmetry of the total action, as is demonstrated by example 1 
in section 4. Our example 2 in that section illustrated that our algorithm 
generates correctly the symmetries of the total action away from the constrained surface, although the structure functions may be singular on that surface, provided the gauge identities are well defined everywhere. However, for the Hamilton equations of motion to exhibit this symmetry, the structure functions must themselves be well defined on the constrained surface. Finally, example 3 in section 4 illustrated that in the case of bifurcations of constraints, the choice of a particular branch may change the structure of the first class constraints, such as to be in conflict with Dirac's conjecture.
We emphasize that such a procedure is excluded in our algorithm.


\begin{thebibliography}{14}
\bibitem{1}
L. Castellani, Ann. Phys. {\bf 143} (1982) 357.
\bibitem{Girotti}
M.E.V. Costa, H.O. Girotti and T.J.M. Simoes, Phys. Rev. {\bf D32} (1985) 405.
\bibitem{3}
X. Gracia and J.M. Pons, Ann. Phys. {\bf 187} (1988) 355.
\bibitem{4}
J. Gomis, M. Henneaux and J.M. Pons, Class. Quantum Grav. {\bf 7} (1990) 1089.
\bibitem{Henneaux}
M. Henneaux, C. Teitelboim and J. Zanelli, Nucl. Phys. {\bf 332} (1990) 169.
\bibitem{BRR} 
R. Banerjee, H.J. Rothe and K.D. Rothe, J. Phys. A: Math. Gen. {\bf 33}
(2000) 2059, and Phys. Lett. {\bf B463} (1999) 248.
\bibitem{Cabo} 
A. Cabo and D. Louis-Martinez, Phys. Rev. {\bf D42} (1990) 2726.
\bibitem{Chaichian}
A. Cabo, M. Chaichian, and D. Louis Martinez, J. Math. Phys. {\bf 34} 
(1993) 5646;  M. Chaichian, D. Louis-Martinez and L. Lusanna,
Annals of Physics (N.Y) {\bf 232} (1994) 40. 
\bibitem{Sudarshan} E.C.G. Sudarshan and N. Mukunda, ``Classical Dynamics. A Modern Perspective",
Wiley, New York (1974); 
A.J. Hanson, T. Regge and C. Teidelboim, ``Constrained Hamiltonian Systems",
Academia Nazional dei Lincei (1976);
D.M. Gitman. and I.V. Tyutin, {\it Quantization of Fields with Constraints}, 
Springer (1980).
\bibitem{Shirzad99}
A. Shirzad and M. Shabani Moghadam, J. Phys. {\bf A}: Math. Gen. {\bf 32} (1999) 8185.
\bibitem{Pyatov} P.N. Pyatov and A.V. Razumov, Int. J. Mod. Phys. {\bf A4} (1989)
3211.
\bibitem{Gomis} J. Gomis, J.Paris and S. Samuel, Phys. Rep. {\bf 259} (1995) 1.
\bibitem{RR} H. Montani, Int. J. Mod. Phys. {\bf 8A} (1993) 4319; 
A. Shirzad, J. Phys. {\bf A}: Math. Gen. {\bf 31} (1998) 2747;
F. Loran and A. Shirzad, Int. J. Mod. Phys. {\bf A17} (2002) 625. 
H.J. Rothe and K.D. Rothe, J. Phys. A: Math. Gen. {\bf 36} (2003) 1671.
\bibitem{Dirac}
P.A.M.Dirac, {\it Lectures on Quantum Mechanics} (Yesiva University Press, 1964).
\bibitem{Heinz}
Heinz J. Rothe, Phys. Lett. {\bf B539} (2002) 296.
\bibitem{Stefano} R. Di Stefano, Phys. Rev. {\bf D27} (1983) 1752.
\bibitem{Cawley} R. Cawley, Phys. Rev. Lett. {\bf 42} (1979) 413; {\it ibid}
Phys. Rev {\bf D21} (1980) 2988.
\bibitem{Teitelboim} M. Henneaux and C. Teitelboim, ``Quantization of gauge systems", Princeton University Press (1992).
\bibitem{Zanelli} O. Miskovic and J. Zanelli, hep-th/0302033.
\bibitem{Lusanna} L. Lusanna, Rivista Nuovo Cimento {\bf 14}, n.3 (1991).
\bibitem{Christ-Lee} N.H. Christ and T.D. Lee, Phys. Rev. {\bf D22} (1980) 939.
\end{thebibliography}
\end{document}